\documentclass[oneside,english,preprint]{elsart}
\usepackage[T1]{fontenc}
\usepackage[latin9]{inputenc}
\usepackage{graphicx}
\usepackage{babel}

\begin{document}
\begin{frontmatter}

\title{Comments on {}``Modelling the gap size distribution of parked cars''}

\author{M. Girardi}

\ead{mauricio.girardi@unipampa.edu.br}

\address{Universidade Federal do Pampa - Caixa Postal 7, 96412-420, Bagé,
Rio Grande do Sul, Brazil.}

\begin{abstract}
In this Comment we discuss some points concerning the modeling of
parked cars proposed in the article by Rawal and Rodgers, Physica
A (2005). We also introduce another approach to this problem which
leads to a better description of the empirical data collected by the
authors.

\end{abstract}
\end{frontmatter}
Rawal and Rodgers \cite{key-1} presented in their article two different
models to describe parked cars in a central area of London. They measured
the distances (gaps) between about 500 cars in four streets, and observed
that the gap distribution have minima for small and large gaps. Figure
1 exhibits a reproduction of the collected data.

In the first proposed model (model A), cars of length 1 are placed
randomly on an infinite linear road if the vacancy is greater than
or equal to $1+\varepsilon$, with $\varepsilon\ge0$. The extra space
$\varepsilon$ mimics the need to manoeuvre. For the full car road,
the gap distribution which they found has a plateau for gaps lower
than $\varepsilon$, decreasing monotonically for gaps of sizes lower
then $1+\varepsilon$. These results are very different from the empirical
one.

In the second model (model B), a car of length 1 can park and stay
in its position with probability $p$, and park, driving to the nearest
car, with probability $1-p$. In this case, the distance $y$ from
the nearest car is chosen with probability $f(y)$. Two functional
forms for $f(y)$ were tested to reproduce the empirical data, and
the authors argued that $f(y)=6y(1-y)$ fits the data very well.

Three points about the above article deserve some discussion:

\begin{enumerate}
\item In their first model, the idea to introduce of an extra space to manoeuvre
was good, and seems to be an important ingredient to any model for
parked cars. However, it was not included in their second model.
\item In model B, the functional form of $f(y)$ had no justification or
explanation, and the authors gave no argument to why they used it.
\item The results obtained from the second model are qualitatively different
from the empirical ones, since the gap distribution does not have
the convex curvature for large gaps.
\end{enumerate}
In this way, we propose a model where the extra space $\varepsilon$
is maintained, while another dynamics for car parking is chosen. Here,
a car is displaced between the two other cars with a Gaussian probability
$p(x)\propto\exp\left(-\frac{(x-x_{0})^{2}}{\sigma^{2}}\right)$,
where $x$ is the distance from the center $x_{0}$ of the vacancy.
This dynamics resembles the drivers desire to leave a secure gap between
front and rear cars (the maximum probability is when the car stays
at the center of the vacancy), and their ability to measure this distance,
given by the standard deviation $\sigma$. 

Figure 2 shows simulation results \cite{key-2,key-3} of the present
model for various values of $\varepsilon$ and $\sigma$. In figure
2a, $\sigma$ is equal to $gap/10$, indicating the case where drivers
are very skillful, and the cars are placed almost at center of the
gap. The opposite case ($\sigma=gap/0.1$) is shown in figure 2d,
where the cars are placed at random in the gap, as in model A of ref.
{[}1]. As can be seen, changing the value of $\varepsilon$ does not
modify qualitatively the gap distribution, except for $\varepsilon=0$
where the distribution is monotonically decreasing. The interesting
results are exhibited in figures 2b and 2c for $\sigma=gap/4$ and
$\sigma=gap/2$ respectively. Here the curves resemble the empirical
data, with lower densities for small gaps (non zero densities), a
maximum at a typical gap size, and a convex curvature of the distribution
for large gaps. This contrasts with Fig. 9 of ref. {[}1], which is
concave for large gaps.

Concluding, we found a model for parked cars which agrees qualitatively
with empirical data for full car roads. The essential ingredients
are the early employed extra space for manoeuvre, and a dynamics where
the driver$^{\prime}$s skill to park his car keeping equal gaps from
rear and front cars is represented by the standard deviation in the
Gaussian probability to put the car in the center of the vacancy.

\begin{ack}
We would like to thank the financial support of the Brazilian agencies
FAPESP and CNPq.
\end{ack}

\section*{Figures}

1 - Gap distribution between parked cars in the central area of London
(same as Fig1. of ref. \cite{key-1}).

2 - Gap size distribution for (a) $\sigma=gap/10$, (b) $\sigma=gap/4$,
(c) $\sigma=gap/2$ and $\sigma=gap/0.1$. In each plot, from left
to right, we have $\varepsilon=0,\,1,\,3,\,5,\,7$ and $9$.

\newpage{}%
\begin{figure}
\includegraphics[bb=0bp 0bp 400bp 400bp,clip]{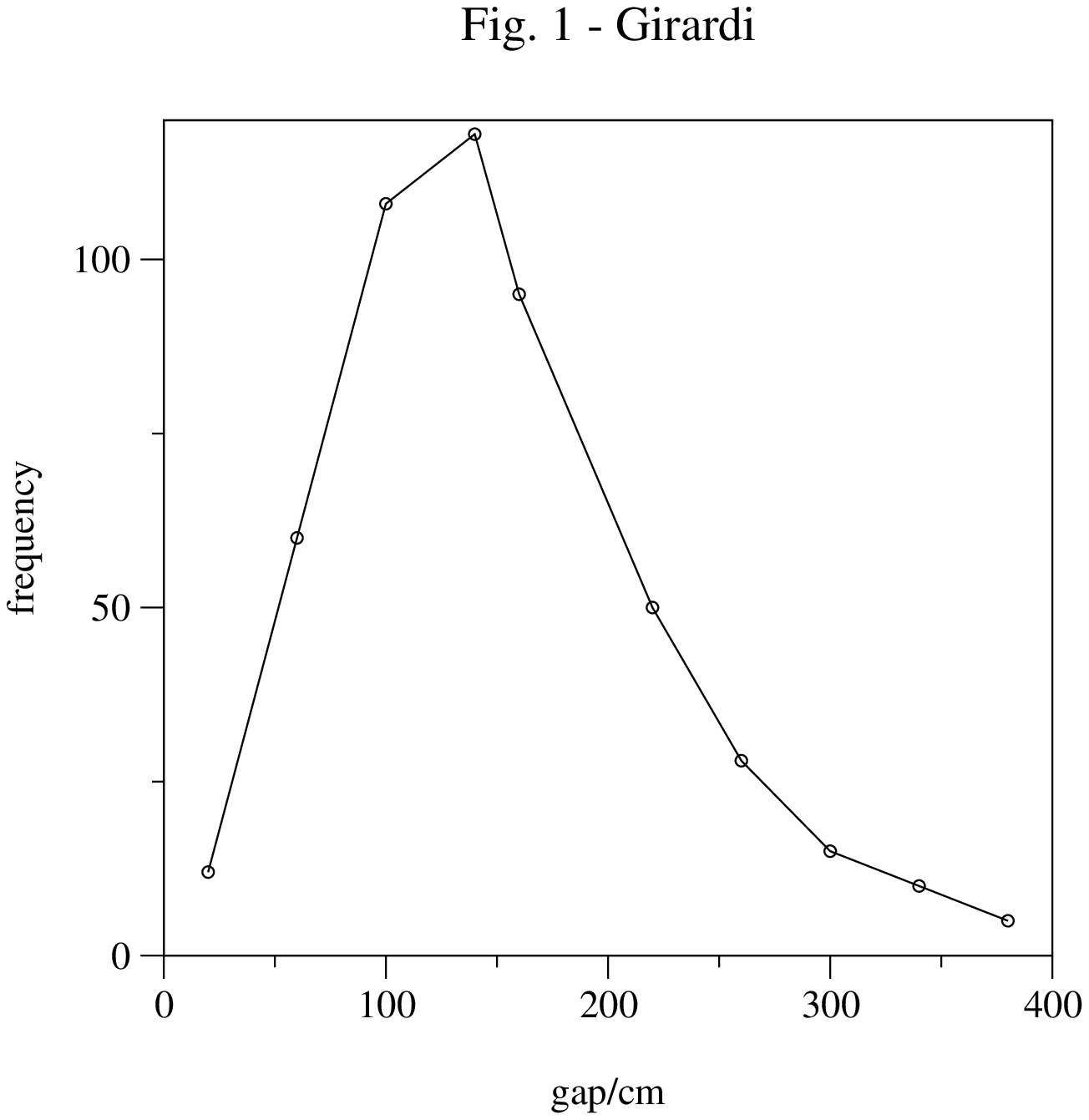}

\end{figure}

\newpage{}%
\begin{figure}
\includegraphics[bb=0bp 0bp 400bp 400bp,clip]{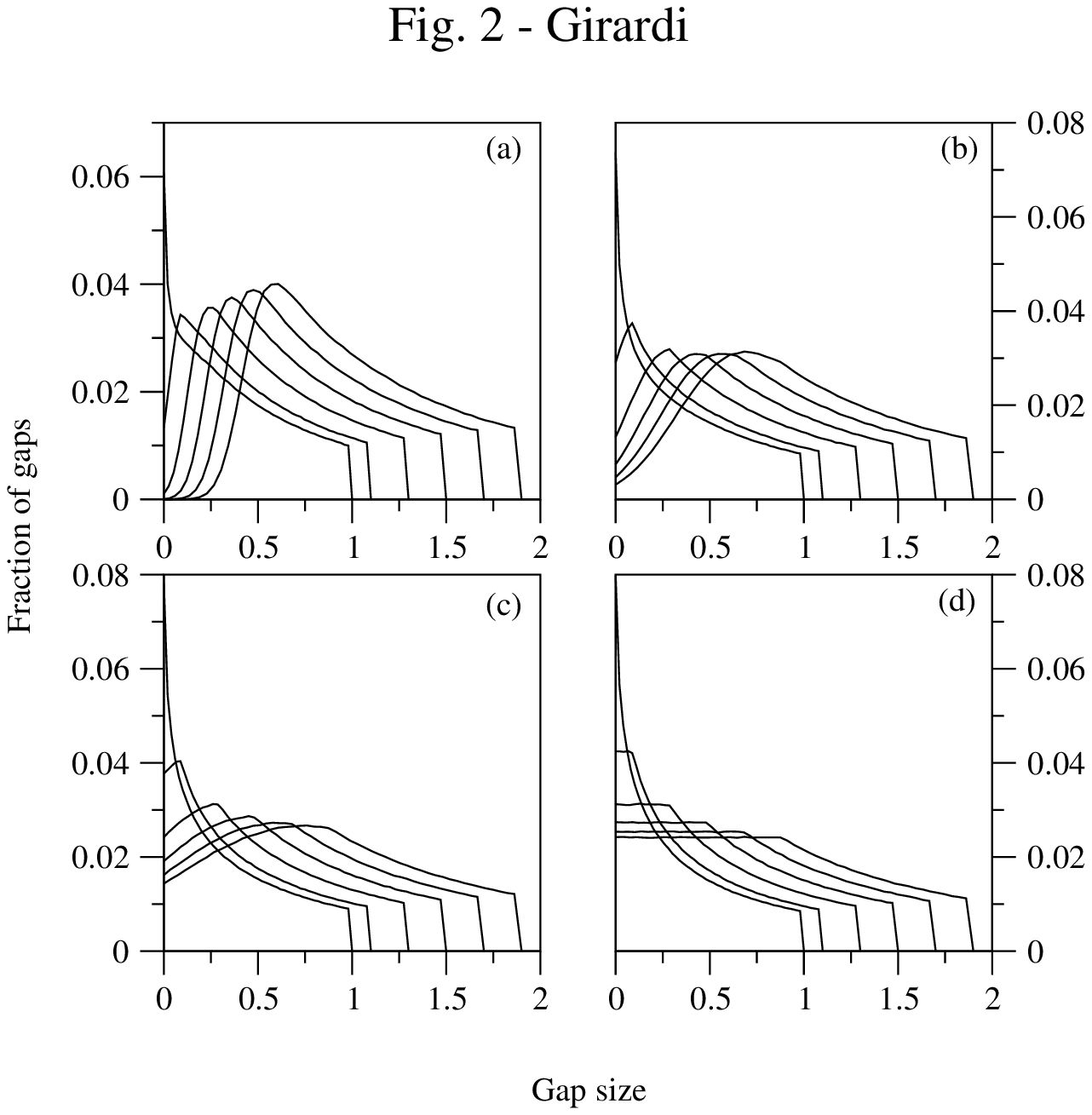}
\end{figure}

\end{document}